# The Inverted Big-Bang

*Our universe appears to have been created not out of nothing
but from a strange space-time dust of quantum geometry*

By Rüdiger Vaas

*Summary:*
• *Quantum geometry makes it possible to avoid the ominous beginning of our universe with its physically unrealistic – infinite – curvature, extreme temperature, and energy density. It could be the long sought after explanation of the big-bang.*
• *It perhaps even opens a window into a time before the big-bang – space itself may have come from an earlier collapsing universe that turned inside out or inverted and began to expand again.*

With the help of one equation, Martin Bojowald tries to look into a time that no one has ever seen – into a time before time, into the time before the big-bang. If this equation is correct, the big-bang was not the beginning of everything but merely a transition – the end of a previous universe collapsing into itself and at the same time turning inside out into a new universe expanding out. The young physicist at Max-Planck-Institute for gravitational physics in Potsdam cannot yet say what happened exactly before the big-bang. But his results are already so promising, that he has received high recognition and collaboration from renowned physicists worldwide.

Obviously equations are not telescopes or time machines that permit us to really peer into this presumed precursor-universe. Yet mathematical intuition and physical genius can help us leap into this fantasy. A quantum leap in viewpoint is also urgently needed to solve perhaps the biggest mystery of physics: the origin of the universe. (By the way, one needs to literally take a quantum leap. Often the ignorant metaphor for a large breakthrough, here it is the smallest step in nature allowed by the laws of quantum physics.)

Classical physics in the shape of Albert Einstein's theory of general relativity first formulated this riddle physically. But it could not solve the problem – rather it pronounced, as it were, its own bankruptcy. Soon after the Russian physicist Alexander Friedmann formulated, in 1922 and 1924, two equations for the evolution of space in a highly simplified form within the framework of relativity theory, it shocked the science with its prediction of the big-bang singularity. General relativity theory predicts a boundary at the big-bang, in that the laws of physics break down as does the applicability of the theory: space and time shrink to nothing while the curvature of space, energy density, pressure and temperature on the other hand grow tremendously. Stephen Hawking and Roger Penrose proved rigorously in the 1960s that this conclusion cannot be avoided in general relativity.

"General relativity predicts a first moment of time," comments Lee Smolin, physics professor at the Canadian University of Waterloo and the associated Perimeter Institute. "But this conclusion disregards quantum physics. For relativity theory is not a quantum theory" and thus, in the last few years the hopes of cosmologists to crack the mystery of the big-bang singularity have grown stronger.

With quantum geometry, a new highly developed tool, physicists now stand to close this painful gap in relativity – that is to say, to fill the gap with a new worldview. For in quantum geometry, space and time are not continuous and flowing, as in relativity (as also assumed in quantum theory), but rather granular and discrete, as if it were made out of space-time atoms. In quantum geometry, this space-time dust is called a spin network or spin foam – the submicroscopic tissue of the world. This tissue, Martin Bojowald and his colleagues believe, has no holes in the big-bang either.

"Quantum physics does not stop at the big-bang," says Abhay Ashtekar, physics professor at Pennsylvania State University and co-founder of quantum geometry. "Classical space-time 'dissolves' near the big-bang, but the spin network is still there." It is to a certain extent eternal. "There was thus no emergence of the universe from 'nothing' because 'nothing' simply does not exist. There was always something already."

In this manner quantum geometry has the philosophical advantage of simply getting rid of apparently unsolvable questions. Here its strength, the independence of the theory from a background space-time metric, makes itself especially noticeable. Ashtekar: "Matter and geometry should both be born together quantum mechanically."

When he was working as a postdoctoral researcher with Ashtekar, Martin Bojowald demonstrated how the spin network could have ignited the big-bang. His quantum-cosmological time evolution equation – a refined quantization of the Friedmann equation – strains the ordinary imagination and comprehension to the extreme. But it triumphs over the large question mark in relativistic cosmology: the big-bang singularity. For out of the perspective of quantum geometry, classical space-time disappeared when our universe was only $10^{-29}$ centimeters in size. This is not to down-play the big bang. At this moment the curvature of space-time was tremendously large even in the context of quantum geometry cosmology – "about $10^{77}$ times that at the horizon of a solar mass black hole," calculated Ashtekar. "But it does not become infinite. The quantum state of the universe is perfectly well-defined. One can study initial conditions at the big-bang and hope to analyze their ramifications for structure formation in the early universe."

The long-range consequence of this extension is described by Ashtekar and Bojowald with Jerzy Lewandowski of the University of Warsaw in the journal *Advances in Theoretical and Mathematical Physics* as follows: "The question of whether the universe had a beginning at a finite time is now 'transcended'. At first, the answer seems to be 'no' in the sense that the quantum evolution does not stop at the big-bang. However, since space-time geometry 'dissolves' near the big-bang, there is no longer a notion of time, or of 'before' or 'after' in the familiar sense. Therefore, strictly, the question is no longer meaningful. The paradigm has changed and meaningful questions must now be phrased differently, without using notions tied to classical space-times."

This discovery is credited to Bojowald's equation, which is based on earlier work of Thomas Thiemann who was formerly also at the Max-Planck-Institute for gravitational physics and is currently a researcher in Canada with Lee Smolin.

"Formulating the equation did not happen in a flash but rather took almost two years. Everything had to be extracted from the very complicated expressions of the full theory," remembers Bojowald who was born in 1973 and graduated from the Rheinisch-Westfälische Technische Hochschule,

Aachen, with a doctorate in physics. "At first I was looking for solutions of quantum geometry describing rotationally symmetric black holes. Symmetries usually simplify the equations, but this was not yet enough. Then I turned to cosmological models which are even simpler thanks to their homogeneity principle: at large scales the universe is very uniform. However, as a physical application I took this seriously only later since to my mind preceding work on quantum cosmology looked arbitrary."

In contrast to the traditional equation of quantum cosmology – the Wheeler-DeWitt equation – it does not have a continuous time evolution but a discrete one (see the appendix "Quantum cosmology for the curious"). The passage of time proceeds in steps, albeit very tiny steps. The Wheeler-DeWitt equation arises from it as an approximation for larger time intervals and space volumes. But on the Planck-scale in the big-bang, when our observable universe was not much more than a point, Bojowald's equation supplies other new solutions that are physically realistic and contain no singularity. That is an enormous breakthrough.

"There are two sides to the singularity problem," says Bojowald. "First, energy densities become infinite – but this is prevented by quantum geometry as follows from work by Thomas Thiemann. More serious is the break-down of time evolution." This problem has now been solved by Bojowald: "In the context of quantum geometry we can now explicitly study the evolution, and we see that it no longer breaks down. In a sense, we can calculate backwards to times 'before' the big-bang, where 'before' is not to be confused with our usual understanding of time."

The big-bang in this picture corresponds to the minimal expansion of the universe. It amounts to only about 0.36 Planck-lengths, as Bojowald has calculated (a Planck-length is $10^{-35}$ meters). "The value is determined by the square root of the so-called Barbero-Immirzi parameter, which at first is a free parameter of the theory but can be fixed by black hole entropy calculations. This shows a relationship between the different computations." This example also illustrates how theorists check their theories in the absence of observational data: the freedom from contradiction and self-consistency are signs of superior quality. "It is a success that the value is close to the Planck-length. A priori, concrete calculations could have resulted in a value like, say, $10^{20}$ Planck-lengths, which would already be ruled out by accelerator experiments. There is, of course, still a lot of space between 0.36 and $10^{20}$, but cosmological models would already run into technical problems if it would be around 3 rather than 0.36. This gives rise to several consistency checks, which so far have all been passed."

The minimum expansion marks the transition from an earlier phase of the universe. Which characteristics it had – whether it had a classical space-time at all and perhaps matter, or whether it was in a totally crazy quantum state – cannot be read off from the calculations yet. "Our hope is that one day we will be able to restrict the freedom further, for instance, by consistency conditions of quantum geometry or empirical data such as the cosmic background radiation," Bojowald says. It is also not necessary for the precursor-universe to be collapsed as a whole, but rather it is sufficient that a collapsed black hole therein would contain the germ for our universe – and other black holes would bear accordingly different universes. Actually Lee Smolin had already speculated some years ago about such a cosmic genesis. Anyhow, if Martin Bojowald is right, the big-bang was only a transitional phase and not the beginning of everything.

Particularly bizarre is the inversion of space at the big-bang. "Space in a sense turns its inside out," Bojowald says. "This can be visualized with an ideally spherical balloon which looses air. It remains an empty balloon such that all its parts clash together – as in a singularity. Now one has to imagine that instead of clashing the parts can freely move through each other and simply move forward. The balloon then again expands with the former inside pointing outward, and vice versa."

Bojowald's extension has a big advantage in comparison with competing cosmological theories: it suffices to take the characteristics of the universe today as input for the equations and reckon backwards in time. Apart from the well-known natural constants, it is crucial to assume the common experience that nature does not change radically over short time scales. In this quantum cosmological beginning, the initial conditions of the universe become a part of the laws of nature and are thus no longer puzzling and unexplainable.

"That there is no freedom to choose initial or boundary conditions is new to physics and has philosophical consequences," says Bojowald. "Usually, dynamical laws are separate from and independent of boundary conditions. The new way of deriving such conditions comes from the direct picture we have of classical singularities and their replacement by quantum geometry. There is no need to introduce auxiliary constructions which most of the time are not physically sound and so dependent on one's personal taste."

The inflation of the universe – its sudden, exponential swelling shortly after the big-bang – a model which is accepted by many cosmologists today, also has a place in quantum geometry. Bojowald: "You can even regard it as a prediction of quantum geometry. For the first time it could be derived from a theory of quantum gravity." And even the postulate of the yet unknown inflaton-field is not necessary for driving the inflation.

Certainly numerous details are still unsettled, and possibly the quantum geometrical inflation may not be sufficient to allow the universe to become large enough. But even if an additional inflaton-field has to be assumed, its characteristics would be less special than in previous models and can be estimated using quantum geometry. This has observable consequences, as Roy Maartens of the University of Portsmouth and his colleagues Shinji Tsujikawa and Parampreet Singh have recently shown. It could explain even the measured temperature variations in the cosmic background radiation – the afterglow of the big-bang – better than other models. "Already now is it possible to test such models given by quantum geometry," Bojowald is pleased – and adds a matching restriction: "With all that, of course, one has also be careful since there are no really decisive tests yet."

At the end of the inflationary period the universe heated up and matter was formed. All this already happened in the first fraction of a second, after the universe reached its minimum size. With the expansion of space, the spin network continued to grow. And just a few 100000 Planck-times later (a Planck-time is $10^{-43}$ seconds), the continuous fabric of classical space-time was formed. Indeed, the text on this page is woven from the very same fabric: no less than $10^{68}$ quantum threads pass through this sheet of paper alone.

*Appendix*

# Quantum Cosmology for the Curious

*The whole world in one equation: formulas for a universal description*

In his preface to the best seller *A Brief History of Time*, Stephen Hawking worried that each formula in the book would halve the number of its readers. If this also applies to this text, please browse through it quickly – otherwise it promises to threaten the author with more hardships. For his naive impertinence caused him to be bold enough to offer a glimpse of the fundamental equations with which the quantum cosmologists try to describe nothing less than the origin and development of the entire universe – cosmic formulas par excellence indeed.

In quantum cosmology, this happens by means of the Wheeler-DeWitt equation. Bryce DeWitt published it in 1967 in the journal *Physical Reviews* – based on initial work by physics professor John Archibald Wheeler at the University of Texas in Austin. It reads: $\mathbf{H\psi = 0}$.

That sounds short and crisp and can be easily printed on a T-shirt. In parties, such a T-shirt would provide material for conversation. But beware – probing questions are guaranteed! In order not to cause an embarrassing silence, it is worthwhile to delve into a few details.

$\mathbf{H\psi = 0}$ is the Hamiltonian constraint, a fundamental equation, with which the quantum-cosmological theory of the universe is specified. A constraint equation describes the interaction of physical fields – for instance of the force of gravity and matter – by certain functions that are first considered independent from one another. It guarantees that, for example, a change of matter goes with a change in gravitation and vice versa. With $\mathbf{H\psi = 0}$ alone one cannot do much. It is crucial to solve the Wheeler-DeWitt equation. It can be derived from the Hamiltonian constraint equation as follows:

$$-1/6 \cdot l_{Pl}^4 \cdot ((a \cdot \mathbf{\psi}(a, \mathbf{\phi}))' \cdot 1/a)' \cdot 1/a + 3/2 \cdot k \cdot a \cdot \mathbf{\psi}(a, \mathbf{\phi}) = 8\pi G \cdot H_\phi(a)\mathbf{\psi}(a, \mathbf{\phi})$$

That looks complicated and indeed it is. Therefore let us move very slowly step by step:

• $l_{Pl}$ is the Planck-length ($10^{-35}$ meters)

• **a** is the scale factor of the universe. It describes the variation in the size of space and can be used therefore also as a measure of "internal time" of the universe. If space does not expand uniformly, but rather anisotropically – for example into one direction more quickly than in another – one must use different, direction dependent sizes $\mathbf{a_1}$, $\mathbf{a_2}$ etc.

• $\mathbf{\psi}(a, \mathbf{\phi})$ is the wave function of the universe – to a certain extent the universal extension of the Schrödinger-equation basic in quantum theory, applied now to the universe as a whole. $\mathbf{\psi}$ depends on the scale factor **a** (the stroke **'** in the equation stands for the mathematical derivative – it is therefore a differential equation) as well as on the nature of matter $\mathbf{\phi}$.

- The Hamiltonian operator **H** (named after the Scottish physicist and mathematician William Rowan Hamilton) is an especially tricky instrument used by physicists. In quantum physics it designates an energy operator, which plays a decisive role in the description of the dynamics and interaction of a quantum system. While in classical physics the energy can be calculated directly from the field values, in quantum physics it must be extracted from the wave function $\psi$. Mathematically this happens through the Hamiltonian operator, which one must apply, for example using derivatives, on the wave function.

- $H_\phi(a)$ is the Hamiltonian operator for matter. It contains the entire theory of matter, more exactly: the total energy of all matter fields. (The Hamiltonian operator could in principle also contain the matter description of string theory.) "Matter" is therefore a broad concept that includes not only all the well-known elementary particles as well as radiation and gravitational waves, but also the ominous dark matter that does not interact electromagnetically, fields like the hypothetical inflaton that shaped the epoch of cosmic inflation, and the mysterious dark energy that accelerates the expansion of the universe.

- **k** designates geometry or global curvature of space (0 = flat, +1 = positively curved like a sphere, -1 = negatively curved like a saddle).

- **G** stands for Newton's gravitational constant ($6.6742 \times 10^{-11}$ cubic meter per kilogram seconds square).

If one sets $a = 0$, the differential equation becomes singular. This point corresponds to the classical big-bang singularity where space and time disappear and energy, matter density and temperature become infinite. The earlier attempts of quantum cosmologists – for instance of Bryce DeWitt, Stephen Hawking, Alexander Vilenkin and Andrei Linde – tried to select the initial conditions of the Wheeler-DeWitt equation, as it were, by hand or to extend the equation in such a way that the singularity disappears or the conditions for $a = 0$ do not become unphysical. But that seems artificial and cannot be established through more basic principles.

From the perspective of quantum geometry, space and time in the Wheeler-DeWitt equation are still regarded as continuous. When one gives up this condition, the world looks completely different – and the singularity problem disappears. This has far reaching consequences (see the main article).

Martin Bojowald of the Max-Planck-Institute for gravitational physics in Potsdam succeeded in making this quantum cosmological breakthrough – a thrust into the time before the big-bang. It was achieved in the framework of quantum geometry, defusing the problems of the Wheeler-DeWitt equation by replacing it with a new equation. At the moment of the big-bang the equation leads to other solutions, but at later times it is in excellent agreement with the solutions of the Wheeler-DeWitt equation – which is therefore still useful.

Bojowald's time development equation reads as follows:

$$(V_{|n+4|/2} - V_{|n+4|/2 - 1}) \cdot e^{ik} \cdot \psi_{n+4}(\phi) - (2 + \gamma^2 k^2) \cdot (V_{|n|/2} - V_{|n|/2-1}) \cdot \psi_n(\phi)$$
$$+ (V_{|n-4|/2} - V_{|n-4|/2 -1}) \cdot e^{-ik} \cdot \psi_{n-4}(\phi) = - 8\pi/3 \cdot G \cdot \gamma^3 \cdot l_{Pl}^2 \cdot H_\phi(n) \psi_n(\phi)$$

- **V** is the volume of the universe and **k** again its geometry (here only applicable for 0 or +1). In this equation, the volume cannot approach zero as closely as desired, but rather is bounded by the Planck-length $l_{Pl}$. The minimal volume is:

$$V_{min} = (1/12\sqrt{2}) \cdot \gamma \cdot \sqrt{\gamma} \cdot l_{Pl}^3$$

- **i** is the imaginary number, defined by $i^2 = -1$

- **e** is Euler's number (2.71828182845905), the basis for natural logarithm; additionally, it satisfies: $e^{i\pi} = -1$

- **γ** is the Barbero-Immirzi parameter. Named after the Spanish physicist Fernando Barbero and the Italian physicist Giorgio Immirzi this number indicates the discreteness of the area in quantum geometry in relation to the Planck scale, given by $\sqrt{\gamma} \cdot l_{Pl} = 0.36 \, l_{Pl}$. This value was determined from the calculation of black hole entropy by Abhay Ashtekar, John Baez, Alejandro Corichi and Kirill Krasnov.

- **n** is the crucial part of the equation: it is integer valued, and thus signifies jumps in the progress of time and replaces to a certain extent the scale factor. Negative values of **n** designate the time before the big-bang, positive **n** the time after. The formula takes the absolute value of **n** in the formula for **V**, so that the volume of the universe is always positive. The volume is minimal with **n** = -1 and **n** = 1. (With **n** = 0 the wave function is decoupled and does not have a fixed value – it is still unclear what this means exactly).

In contrast to the Wheeler-DeWitt equation Bojowald's equation is not a differential equation but rather a difference equation. The reason for it is the discrete internal time structure. That means: The time **n** does not "flow" continuously (and thus does not have arbitrary values represented by real numbers), but rather runs in abrupt steps (represented by integers). It is similar to a movie film whose frames follow in such quick succession that our eyes believe and the brain recognizes it as one fluent movement – yet in reality it is a rapid sequence of single snapshots. Today – and already fractions of a second after the big-bang – the film of the universe runs in such a way that it can be described very well with classical or semi-classical physics (the Friedmann equation in general relativity and the Wheeler-DeWitt equation). The individual snapshots of the big-bang become recognizable only under the sharpened eye of quantum geometry.

---

**Acknowledgements:**
This article is the third in a row. It is a pleasure to thank Abhay Ashtekar, Martin Bojowald, and Amitabha Sen again for their excellent collaboration. Without them, this trilogy on quantum geometry would not have been possible both in German and in English.

**Contact:**
Ruediger.Vaas@t-online.de

*Further reading*

Two introductions:

• A popular account of quantum geometry (loop quantum gravity) for the general audience:
Rüdiger Vaas: *Jenseits von Raum und Zeit.* bild der wissenschaft, no. 12 (2003), pp. 50–56.
English translation by Amitabha Sen:
*Beyond Space And Time*
http://arxiv.org/abs/physics/0401128

• About the controversy between quantum geometry and string theory:
Rüdiger Vaas: *Das Duell: Strings gegen Schleifen.* bild der wissenschaft, no. 4 (2004), pp. 44–49.
English translation by Martin Bojowald and Amitabha Sen:
*The Duel: Strings versus Loops*
http://arxiv.org/abs/physics/0403112

Both articles are also available online both in German and English via
http://cgpg.gravity.psu.edu/research/poparticle.shtml
and
http://cgpg.gravity.psu.edu/people/Ashtekar/articles.html

For a review of the scientific details see:

• Martin Bojowald: *Loop Quantum Cosmology: Recent Progress.* Plenary talk at ICGC 04, Cochin, India.
http://arxiv.org/abs/gr-qc/0402053
• Martin Bojowald, Hugo A. Morales-Tecotl: *Cosmological applications of loop quantum gravity.*
Lecture Notes in Physics, vol. 646 (2004), pp. 421–462.
http://arxiv.org/abs/gr-qc/0306008
• Abhay Ashtekar, Martin Bojowald, Jerzy Lewandowski: *Mathematical structure of loop quantum cosmology.* Advances in Theoretical and Mathematical Physics, vol. 7 (2003), pp. 233–268.
http://arxiv.org/abs/gr-qc/0304074

---

Translated by Amitabha Sen with permission from
Rüdiger Vaas: *Der umgestülpte Urknall.*
bild der wissenschaft, no. 4 (2004), pp. 50-55.8